\documentclass[fleqn,usenatbib]{mnras}

\usepackage{newtxtext,newtxmath}
\usepackage{mdframed}
\usepackage[T1]{fontenc}
\usepackage{ae,aecompl}
\usepackage{graphicx}	
\usepackage{amsmath}	
\usepackage{amssymb}	
\usepackage{todonotes}
\usepackage[capitalize]{cleveref}
\usepackage{tikz}
\usetikzlibrary{shapes,arrows}

\Crefname{equation}{Equation}{Equations}
\crefname{equation}{Eq.}{Eqs.}
\Crefname{figure}{Figure}{Figures}
\crefname{figure}{Fig.}{Figs.}
\Crefname{table}{Table}{Tables}
\crefname{table}{Tab.}{Tabs.}
\Crefname{section}{Section}{Sections}
\crefname{section}{Sec.}{Secs.}

\def\LCDM{$\Lambda$CDM}
\def\Planck{{\it Planck}}
\def\rmd{\mathrm{d}}
\def\Z{\mathcal{Z}}
\def\P{\mathcal{P}}
\def\L{\mathcal{L}}
\def\D{\mathcal{D}}

\def\C{\Sigma}
\def\E{\Lambda}
\def\Lmax{\L^{\rm max}}
\def\T{\mathsf{T}} 

\title[Quantifying Suspiciousness Within Correlated Data Sets]{Quantifying Suspiciousness Within Correlated Data Sets}

\author[Lemos et al.]{
Pablo Lemos,$^{1}$\thanks{E-mail: pablo.lemos.18@ucl.ac.uk}
Fabian K{\"o}hlinger,$^{2}$
Will Handley,$^{3, 4}$ 
Benjamin Joachimi,$^{1}$
\newauthor{ Lorne Whiteway$^{1}$ and Ofer Lahav$^{1}$}\\
\\
$^{1}$Department of Physics and Astronomy, University College London, Gower Street, London, WC1E 6BT, UK\\
$^{2}$Kavli IPMU (WPI), UTIAS, The University of Tokyo, Kashiwa, Chiba 277-8583, Japan\\
$^{3}$Astrophysics Group, Cavendish Laboratory, J.J.Thomson Avenue, Cambridge, CB3 0HE, UK\\
$^{4}$Kavli Institute for Cosmology, Madingley Road, Cambridge, CB3 0HA, UK\\
}

\date{Accepted XXX. Received YYY; in original form ZZZ}

\pubyear{2019}

\begin{document}
\label{firstpage}
\maketitle

\begin{abstract}
We propose a principled Bayesian method for quantifying tension between correlated datasets with wide uninformative parameter priors. This is achieved by extending the Suspiciousness statistic, which is insensitive to priors. Our method uses global summary statistics, and as such it can be used as a diagnostic for internal consistency. We show how our approach can be combined with methods that use parameter space and data space to identify the existing internal discrepancies. As an example, we use it to test the internal consistency of the KiDS-450 data in 4 photometric redshift bins, and to recover controlled internal discrepancies in simulated KiDS data. We propose this as a diagnostic of internal consistency for present and future cosmological surveys, and as a tension metric for data sets that have non-negligible correlation, such as LSST and {\it Euclid}.
\end{abstract}


\section{Introduction}

Quantifying consistency between different data sets has become one of the main challenges in modern cosmology. With the increasing number of methods and surveys measuring different properties of the Universe, it is crucial to develop appropriate statistical tools to compare and combine these data sets. This is important for two reasons. First, differences in cosmological constraints from different data sets, i.e. `tensions', could indicate unaccounted-for systematic errors in one or both data sets, or could indicate that the underlying theoretical model (for example \LCDM) is not sufficient to explain both data sets. Second, because the combination of different data sets can provide powerful cosmological constraints by breaking degeneracies existing in single data sets, but data sets can only be combined meaningfully if they are in agreement.

Cosmological tensions and their correct quantification are now particularly important, given that while all existing data sets match the \LCDM\ model of cosmology individually, there are two main disagreements between data sets which could be a hint of beyond-\LCDM\ physics. This belief is reinforced by the fact the tensions are between high redshift measurements of the Cosmic Microwave Background (CMB) by the \Planck\ satellite \citep{PlanckParameters2015, PlanckParameters2018}, and low redshift measurements of the Hubble constant \citep{Riess2018, Riess2019} and of the growth of structure measured by galaxy clustering and weak gravitational lensing by several surveys such as the Canada France-Hawaii Telescope Lensing Survey \citep[CFHTLenS,][]{Heymans2012, Heymans2013, Joudaki2017}, the Kilo Degree Survey \citep[KiDS,][]{Hildebrandt2017, Kohlinger2017}, and the Dark Energy Survey \citep[DES,][]{Troxel2017, DES2017}.

There is a recent and extensive literature on quantifying tension in the context of cosmology \citep[for a review of some methods see][]{Charnock2017}. Perhaps the most frequently used is the Bayes ratio $R$ introduced in \cite{Marshall2006}, which has the advantages of using exclusively Bayesian quantities and of being parameterization independent. However, this method has the disadvantage of being proportional to the prior volume (which can hide existing tensions when broad priors are chosen with the goal of being uninformative). Other approaches based on differences in the best fit parameters were introduced by \cite{Lin2017, Lin2017b, Raveri2018, Adhikari2019}. \cite{Kunz2006, Seehars2014, Grandis2016, Grandis2016b, Nicola2019} used methods based on the Bayesian information measured by the Kullback Leibler (KL) Divergence \citep{Kullback1951}. One of the latest methods suggested was the Bayesian `Suspiciousness' introduced in \cite{Handley2019a} (henceforth H19). The Suspiciousness acts as an extension of the $R$ statistic of \cite{Marshall2006} to the case of uninformative priors, preserving many of the desired properties of $R$ without having dependence on prior volume. It is important to point out that both $R$ and $S$ rely on quantities, such as the Bayesian evidence, that are noisy statistics. This was already discussed by \cite{Jenkins2011}, and will be further investigated in \cite{Joachimi2019}.

However, all the methods above assume that the data sets are uncorrelated. In this work, we focus on methods to quantify consistency between correlated data sets (i.e. datasets for which $P(A, B) \neq P(A) P(B)$). The measurement of tension in correlated data sets has numerous applications: it can be used to quantify consistency between different `splits' or parts of a given data set, as a measure of internal consistency. It is also relevant when combining different data sets with non-trivial correlations. For example this will be necessary to obtain combined constraints from Euclid \citep{Laureijs2011} and the Large Synoptic Spectroscopic Survey \citep[LSST,][]{Ivezic2019}, which as showcased by \cite{Rhodes2017, Schuhmann2019} would be very well motivated. 

The problem of assessing consistency between correlated data sets was tackled by \cite{Kohlinger2019} (henceforth K19), in the context of the KiDS-450 data. They use three `tiers' of consistency tests: Bayesian evidence ratios, parameter differences, and posterior predictive distributions (PPD). They conclude that Bayesian evidence ratios are not appropriate for this problem because of their dependence on prior volume. As already mentioned, there are several important advantages to the use of Bayesian evidence ratios with respect to the other two tiers of consistency test introduced in K19. Therefore, in this paper, we extend the work of H19 and apply the Suspiciousness statistic to the case of correlated data sets with broad non-informative priors. This provides a Bayesian, parameterisation-independent measure of consistency between correlated data sets. 

In \cref{sec:theory} we describe how the Suspiciousness can be extended to correlated data sets. \cref{sec:gaussian} applies this method to a toy model consisting of Gaussian data sets that we `split' in two. We discuss how this method is related to the other two tiers of internal consistency introduced in K19 in \cref{sec:ppd}. The method is applied to the KiDS-450 data in \cref{sec:data}. We present our conclusions in \cref{sec:conc}.

\section{Tension in correlated datasets}
\label{sec:theory}
The goal of this section is to extend the tension metrics and Bayesian model dimensionality introduced in \cite{Handley2019a, Handley2019b} to the case of correlated data sets. To derive the $R$ statistic, \cite{Marshall2006} propose comparing two hypotheses

\begin{itemize}
\item $H_0$: There exists one common set of parameters that describes both data sets.
\item $H_1$: There exist two sets of parameters, one for each data set.
\end{itemize}
Bayes' theorem is then used to calculate the ratio of degrees of belief in each hypothesis in light of the data $D$
\begin{equation}
    { P(H_0 | D) \over P(H_1 | D)} = {P(D | H_0) \over P(D |H_1)} \cdot {P(H_0) \over P(H_1)}\propto  { \Z_0 \over \Z_1} = R,
    \label{r01}
\end{equation}
where $R$ is the Bayes ratio and $\Z$ is the Bayesian evidence
\begin{equation}
\label{Z}
\Z_i \equiv P(D|H_i) = \int \rmd\theta \L(\theta) \Pi (\theta),
\end{equation}
with $\theta$ the parameters of the model, $\L(\theta) \equiv P(D | \theta, H)$ the likelihood, and $\Pi (\theta) \equiv P(\theta | H)$ the prior. 
If our prior belief in both hypotheses is the same $P(H_0) = P(H_1)$, the proportionality constant in \cref{r01} becomes unity. Under this formulation we may view tension quantification as a model comparison problem with $R$ as a figure of merit.   
Finally, the posterior from Bayes theorem and the KL Divergence \citep{Kullback1951} are defined via
\begin{eqnarray}
    P(\theta|D,H) \equiv \P(\theta) = \frac{\L(\theta)\Pi(\theta)}{\Z}, \\
    \label{KL}
    \D = \int \rmd \theta \ \P \log \left( \frac{\P}{\Pi} \right).
\end{eqnarray} 

The Bayes ratio approach using $R$ to quantify tension for correlated data sets was used in K19, but, as they show, $R$ depends on the prior volume. This is the same prior volume dependence that is extensively discussed in H19 for uncorrelated data sets. While this would not be a problem for well-motivated priors, it means that we cannot rely on Bayesian evidence ratios obtained with priors that are purposefully chosen to be broad (with the intention of being uninformative).

To address these concerns about prior volume, we can instead use the Suspiciousness $S$ introduced in H19, which can be understood as the value of $R$ that corresponds to the narrowest possible priors that do not significantly alter the shape of the posteriors. The natural logarithm of the Suspiciousness is given by
\begin{equation}
\log S = \log R - \log I.
\label{eqn:S}
\end{equation}
The information $I$ quantifies the {\em a-priori\/} probability that the data sets would match given the prior range. The larger the prior range relative to the posterior constraints, the lower the probability that the constraints will be consistent. For correlated data sets the natural logarithm of the information is given by
\begin{equation}
\label{I10}
\log I = \D_1 - \D_{0}, 
\end{equation}
In the uncorrelated case, the additivity of the KL divergence implies that $\D_1 = \D(A) + \D(B)$ and $\D_0=\D(A,B)$, which recovers the methodology of H19.


\section{Gaussian Example}
\label{sec:gaussian}

\begin{figure}
    \includegraphics[width=0.49\textwidth]{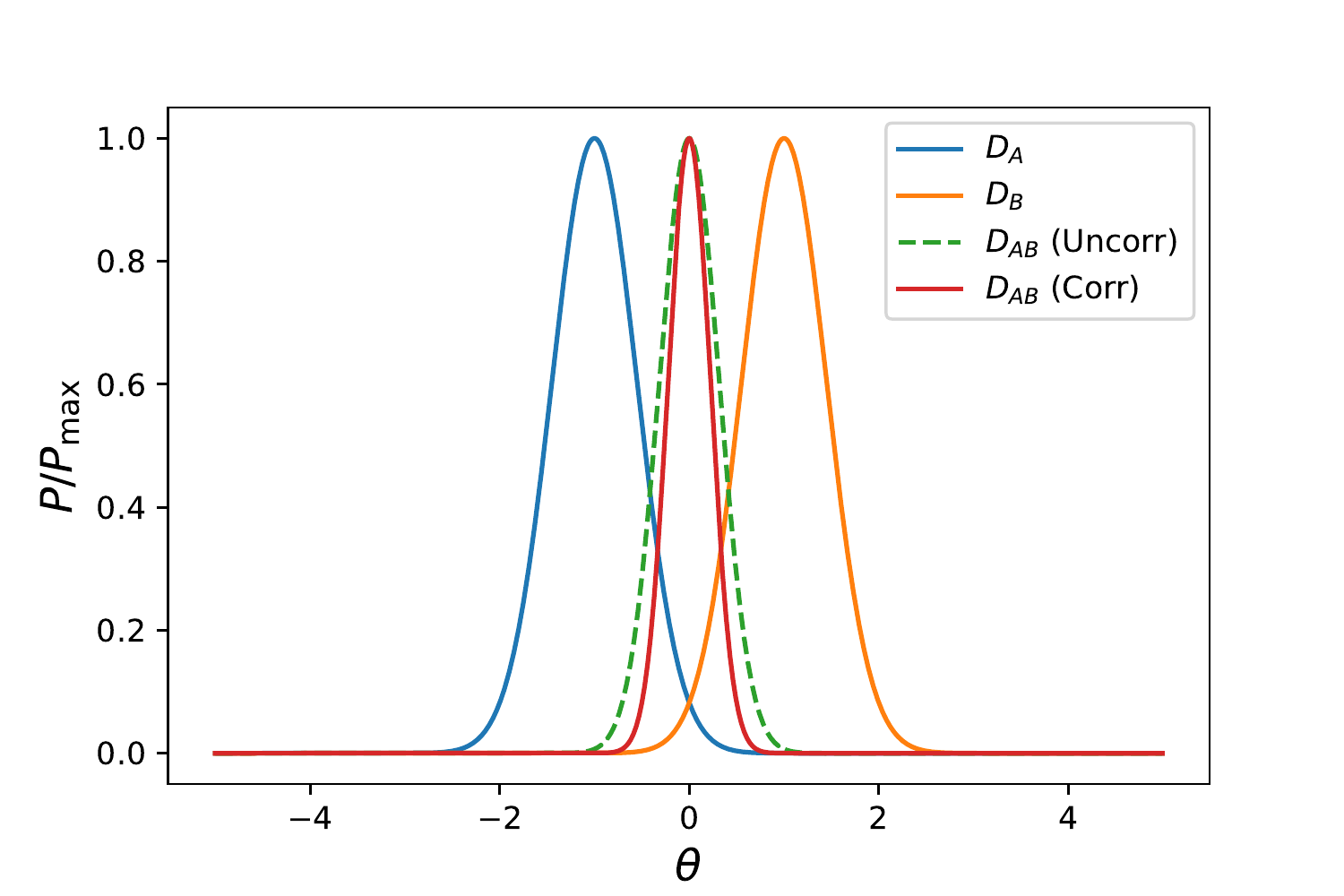}
    \caption{Graphical illustration of the Gaussian example, in the case of a single parameter. The plot shows the two posteriors, and their combination in the uncorrelated and correlated case, for $\sigma_A^2 = \sigma_B^2 = 0.1$, and $V_{\pi} = 10$.
\label{fig:gaussians}}
\end{figure}

To illustrate the formalism, we consider the example of multivariate Gaussian posterior distributions (\cref{fig:gaussians}). Let $A$ and $B$ be two data sets that can constrain the same set $\theta$ of $d$ parameters. Each data set $A$ or $B$ individually constrains one set of parameters $\theta_A$ or $\theta_B$ giving a posterior with parameter mean and covariance of $(\mu_A,\Sigma_A)$ or $(\mu_B,\Sigma_B)$ respectively.

Assume now that the two data sets are correlated. If we combine the two data sets and use hypothesis $H_1$, then the mean and covariance of the marginal distribution of each individual set of parameters will be unchanged; however the data correlation will induce a correlation between parameter sets so that the likelihood will be
\begin{equation}
\log \L_1 = \log \Lmax_1 - {1 \over 2} {\begin{bmatrix}\theta_A-\mu_A\\ \theta_B-\mu_B\end{bmatrix}}^\T \:\begin{bmatrix} \Sigma_A & \Sigma_X \\ \Sigma_X^\T & \Sigma_B \end{bmatrix}^{-1}\: {\begin{bmatrix}\theta_A-\mu_A\\ \theta_B-\mu_B\end{bmatrix}},
\label{eq:corr}
\end{equation}
where the cross matrix $\Sigma_X$ will be zero in the uncorrelated case.

The likelihood for hypothesis $H_0$ can be found by setting ${\theta_A=\theta_B=\theta}$ in \cref{eq:corr}. Notationally it is helpful to define the precision matrix $\Lambda$ in block form via
\begin{equation}
    \C_1 = \begin{bmatrix} \Sigma_A & \Sigma_X \\ \Sigma_X^\T & \Sigma_B \end{bmatrix},\qquad
    \C^{-1}_1 = \E = \begin{bmatrix} \E_A & \E_X \\ \E_X^\T & \E_B \end{bmatrix},
\end{equation}  
where the different blocks in $\Lambda$ are related to those in $\C_1$ by Schur's complement. \citep{Woodbury1950, Zhang2005}.  After some effort manipulating matrix expressions (see Appendix), we find
\begin{align}
\label{eq:sep}
\log \L_0 &= \log \Lmax_0 - {1 \over 2} (\theta - \mu_0)^\T \C_0^{-1} (\theta - \mu_0),\\
\label{eq:system0}
\log \Lmax_0 &= \log \Lmax_1 -{1 \over 2} (\mu_A-\mu_B)^\T \Sigma_{\Delta\mu}^{-1} (\mu_A-\mu_B), \\
\Sigma_{\Delta\mu} &=  \Sigma_A - \Sigma_X-\Sigma_X^\T + \Sigma_B ,\\ 
\mu_0 &= \C_0 \left[ (\E_A +  \E_X^\T) \mu_A + (\E_X + \E_B) \mu_B \right],\\
\label{eq:system3}
\C_0 &= (\E_A +  \E_X + \E_X^\T+ \E_B)^{-1}.
\end{align}
As discussed in H19, the evidence and KL divergence for a multivariate Gaussian with a flat prior of volume $V_\Pi$ that essentially completely encloses the posterior are
\begin{align}
\log \Z &= \log \Lmax + {1 \over 2} \log \left| 2 \pi \C \right| - \log V_{\Pi},
\label{gauss_ev}\\
\D &= \log V_{\Pi} - {1 \over 2} (d +  \log \left| 2 \pi \Sigma \right| ). 
\label{gauss_KL}
\end{align}
Combining \cref{r01,I10,eqn:S,eq:corr,eq:sep,eq:system0,gauss_KL,gauss_ev} yields
\begin{align} 
    \log R &= \log V_{\Pi} + \log\Lmax_0 - \log\Lmax_1 + {1 \over 2} \log \frac{\left| 2 \pi \C_0 \right|}{\left| 2 \pi \C_1 \right|},\\
    \log I &= \log V_{\Pi} - {d \over 2}  + {1 \over 2} \log \frac{\left| 2 \pi \C_0 \right|}{\left| 2 \pi \C_1 \right|},\\
    \log S &= {d \over 2} -{1 \over 2} (\mu_A-\mu_B)^\T \Sigma_{\Delta\mu}^{-1}(\mu_A-\mu_B). 
\end{align}
Note that a factor $V_\Pi$ emerges in $R$ since $H_1$ has twice as many parameters as $H_0$, but that this dependency on prior volume is mirrored in $I$ which therefore cancels in $\log S$. It can easily be verified that if the data are uncorrelated ($\Sigma_X = \Lambda_X=0$) then these results agree with those of H19.

Now under hypothesis $H_0$, $\mu_A - \mu_B$ will be distributed with mean zero and covariance $\Sigma_{\Delta\mu}$, and thus $(\mu_A-\mu_B)^\T \Sigma_{\Delta\mu}^{-1} (\mu_A-\mu_B)$ will have a $\chi^2_d$ distribution (as can be shown by Cholesky decomposing the covariance matrix). Thus $d - 2 \log S$ must follow a $\chi^2_d$ distribution as well (as was the case in H19). The probability $p_t$ of the data sets being discordant by chance is 
\begin{equation}
\label{pt}
    p_t = \int\limits_{d-2\log S}^\infty \chi^2_d(x) \ \rmd x= \int\limits_{d-2\log S}^\infty \frac{x^{d/2-1}e^{-x/2}}{2^{d/2}\Gamma(d/2)} \ \rmd x.
\end{equation} 

The effective number of dimensions constrained by both data sets $d$ is given by the Bayesian Model Dimensionality (BMD) introduced in \cite{Handley2019b}. In particular, \cite{Handley2019b} show that $d$ is exactly the same as the number of dimensions for a Gaussian likelihood. For correlated data sets, $d$ is given by 
\begin{equation}
\label{d_corr}
d = d_1 - d_0,
\end{equation}
where the number of parameters constrained simultaneously by both datasets is given by subtracting the number of parameters constrained by the combination of both from the number of parameters constrained by each dataset separately, in a similar manner to \cite{Handley2019b} for uncorrelated datasets.

While the results of this section have been obtained for Gaussian likelihoods, they can be applied to more general posteriors. This claim is supported by the fact that the Suspiciousness is invariant under coordinate transformations, such as Box-Cox transformations \citep{BoxCox, Joachimi2011, hiranya_box_cox} which “Gaussianise" the posterior. Furthermore, H19 showed that the Suspiciousness recovers the intuitively correct answers for cosmological examples, which are a somewhat non-Gaussian, particularly in the nuisance parameters.

\section{Connection to other Internal Consistency Tests}
\label{sec:ppd} 

\begin{figure}
    \includegraphics[width=0.49\textwidth]{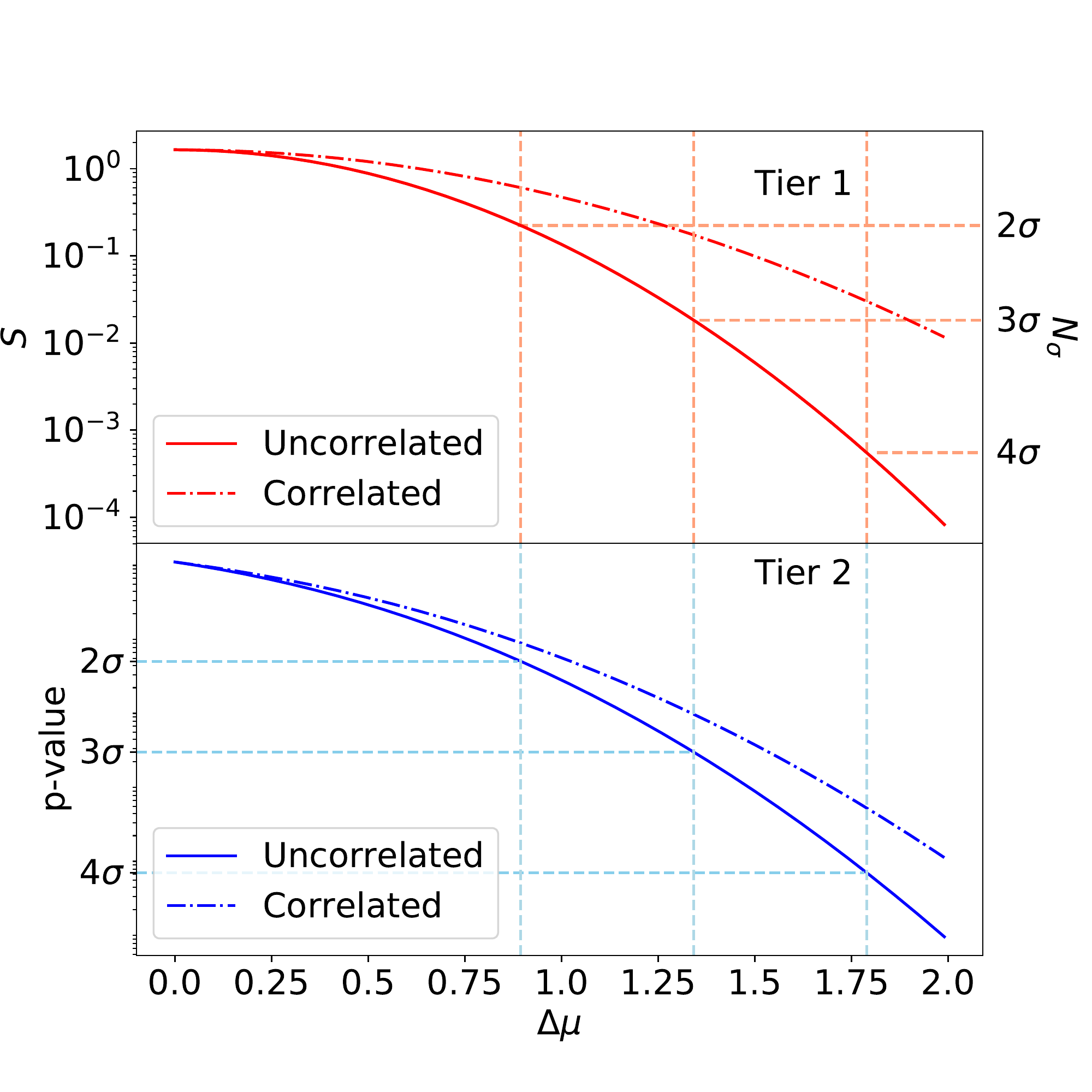}
    \caption{Comparison of tiers one and two of statistical consistency for the toy model of \cref{fig:gaussians}. The top panel shows the Suspiciousness, and the bottom panel shows the p-value obtained from tier two of K19 (parameter differences). The solid curves correspond to uncorrelated Gaussians, while the dash-dotted curve corresponds to the correlated case. The dashed horizontal and vertical lines show the $2, 3, \text{and } 4\sigma$ values for the uncorrelated cases, which agree exactly in this case.  
\label{fig:tiers}}
\end{figure}

K19 proposed three tiers of internal consistency between data sets. We have previously discussed how the Bayesian Suspiciousness serves as an alternative to tier one (the Bayes ratio). In this section we discuss the connection between Suspiciousness and tiers two and three. We argue that, while these three methods are different in their implementation, the underlying quantities being calculated are surprisingly similar. 

Tier two of K19 are differences of parameter duplicates: after calculating the posterior distributions under hypothesis $H_1$ (the data sets are described by different sets of parameters), we can derive the posterior distribution for the differences between the parameters describing each data set. K19 then proposes finding the fraction of samples with a value of the posterior smaller than the value in the origin (which is the point of perfect agreement). This can be seen as an extension of tension metrics based on parameter shifts, such as those introduced in \cite{Raveri2018, Adhikari2019}, to the case of correlated data sets. From the result of \cref{sec:gaussian} it is easy to see how our method is connected to this. For Gaussian posterior distributions, we show that $d - 2 \log S$ is $\chi^2$ distributed, from which we can calculate the tension probability. While the methodology is different, the quantity being calculated is the same in the case of this Gaussian toy model. In fact, one could argue that the `$m \sigma$' interpretation presented in K19 should be dependent on dimensionality, which could be calculated using the BMD \citep{Handley2019b}, and then the calculation of `$m \sigma$' would become similar to \cref{pt}. This is shown in \cref{fig:tiers}, where we compare the results of our method and parameter differences for different one dimensional Gaussian distributions, and confirm that the results are the same.  

Tier three of K19 consists of using the PPD. This technique was introduced in \cite{Gelman1996}, and has been applied to cosmology in recent problems \citep[e.g.][]{Feeney2018, DES-5x2}. When using the PPD, we calculate the probability of data $D_A$ conditional on data $D_B$ and the underlying model $M$ in the following way: 
\begin{equation}
\label{ppd} 
P(D_A | D_B, M) = \int \rmd \theta \ P(D_A | \theta, M) P(\theta | D_B, M). 
\end{equation}
In practice this means that from samples of the posterior for $D_B$, we calculate realizations of the likelihood for $D_A$. The most challenging part of using the PPD is calibrating the distribution. $P(D_A | D_B, M)$ is a probability density which is challenging to normalize. The authors of K19, for example, built a `Translated Probability Distribution' from a predicted data vector, with the goal of calibrating the PPD. However, there is an alternative way of interpreting this number, which is by taking the ratio $P(D_A | D_B, M) / P(D_A|M)$. This quantity is unitless, and can therefore be interpreted as a probability ratio. More importantly, as discussed in \cite{Handley2019a}, this ratio is equal to the Bayes ratio $R$. In this way, we can see tiers one and three of K19 as being elements of the same calculation, but using different calibrations of the PPD. If both methods are correct, then they should produce similar results.

\section{KiDS-450 data}
\label{sec:data}

\begin{figure}
    \includegraphics[width=0.49\textwidth]{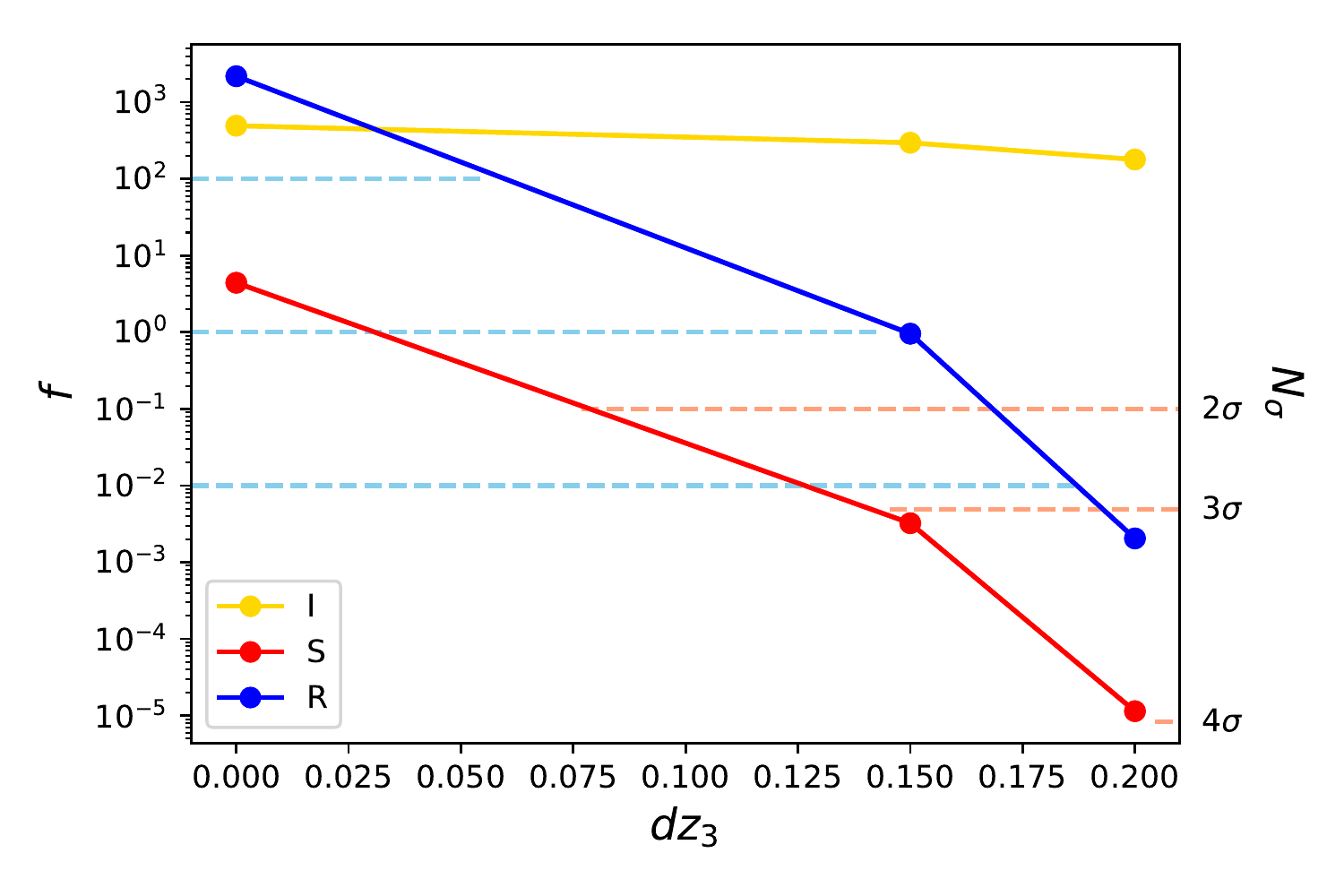}
     \caption{The evidence ratio ($R$, in blue) and the two factors into which it can be split, according to H19: the information ($I$, in yellow) and the Suspiciousness ($S$, in red). The y-axis on the left is the value of each function, which in the case of $R$ corresponds to a ratio of probabilities. The y-axis on the right is the number of sigma corresponding to different values of $S$ (note that the positions of the lines depends on the effective number of constrained directions). The x-axis are different values of $dz_3$ (i.e. shifts in thesource redshift distribution of bin 3) in the sensitivity analysis. The blue line is similar to the left panel in Fig. A1 of K19. 
\label{fig:sens}}
\end{figure}

\begin{table*}
\begin{tabular}{|c | c c | c c c c|}
\hline
$d z_3$ & $\log R$ & Interpretation & $\log S$ & $d$ & $p_t$ & $N_{\sigma}$  \\ 
\hline
$0$ & $7.69 \pm 0.15$  & Decisive Agreement & $1.488 \pm 0.049$ & $2.68 \pm 0.17$ & $0.99995 \pm 0.00053$ & $0$  \\  
$0.15$ & $-0.04 \pm 0.14$  & Neutral & $-5.735 \pm 0.052$ & $2.75 \pm 0.18$ & $0.00203 \pm 0.00028$ & $3.089 \pm 0.041$  \\  
$0.20$ & $-6.18 \pm 0.14$  & Decisive Tension & $-11.383 \pm 0.051$ & $2.60 \pm 0.17$ & $7.6 \cdot 10^{-6} \pm 1.4 \cdot 10^{-6}$ & $4.478 \pm 0.039$  \\  
\hline
\end{tabular}
\caption{Comparison of evidence $\log R$ (interpreted with the Jeffreys scale used in K19) and Suspiciousness $\log S$ for some redshift shifts in the sensitivity analysis. The last three columns show the Bayesian Model Dimensionality $d$, the tension probability $p_t$ and the corresponding number of sigma, calculated as $N_{\sigma} \equiv \sqrt{2} \rm{Erf^{-1}} (1-p_t)$. 
\label{table:sens}} 
\end{table*}

In this section, we test the methods introduced in \cref{sec:theory} on the KiDS-450 data \footnote{See \url{http://kids.strw. leidenuniv.nl/sciencedata.php.}} \citep{Hildebrandt2017}. KiDS uses the correlations in the shapes of galaxy images to measure weak gravitational lensing caused by large scale structure \citep[for reviews, see][]{Bartelmann2001, Kilbinger2015}. Their measurements span a redshift range $z = [0.1, 0.9]$; this range is divided into four redshift bins each of width $\Delta z = 0.2$. The estimators used are the correlation functions $\xi_{+} (i,j), \xi_- (i,j)$, with seven and six angular bins respectively, and where $i, j = 1,..,4$ refer to the redshift bins. The data therefore consist of $130$ datapoints (note that $ \xi_{\pm} (i,j) = \xi_{\pm} (j,i)$). 

We have chosen to examine the KiDS data because \cite{Efstathiou2018} (henceforth E18) found possible inconsistencies between different splits of the data using two simple statistical tests. In particular, bins 3 and 4 were found to be inconsistent with the rest of the data with significances $2.60 \sigma$ and $3.52 \sigma$. Following this work, K19 performed their three tier consistency tests on the same data. Using these statistics, K19 found that the significance of these inconsistencies is reduced, and concluded that the results of E18 are sensitive to the overall goodness of fit of the data. Following E18, \cite{Troxel2018} improved the shot noise model of the analytically derived covariance matrix by including a model for the survey-boundary effects.   This correction increased the size of the error bars at large angular scales, reducing the discrepancies found in E18 to below significant levels. \cite{Asgari2019} also quantified the errors associated with each of the definitions of the angular tomographic bins centre used in \cite{Troxel2018} and \cite{Hildebrandt2017}, proposing an accurate new approach adopted in \cite{Hildebrandt2019}.  In the remainder of this work,  however, we will use the data from \cite{Hildebrandt2017} as analysed by both E18 and K19.
These corrections increased the size of the error bars at large angular scales, reducing the discrepancies found in E18 to below significant levels. However, in the remainder of this work, we will use the data from  \citep{Hildebrandt2017} used by both E18 and K19. 

For our analysis, we compare the KiDS data vectors that involve redshift bin 3 to all others, since bin 3 was one of the discrepant bins in E18. We use the code {\tt 2cosmos}\footnote{\url{https://github.com/fkoehlin/montepython_2cosmos_public}} to sample the KiDS likelihood. {\tt 2cosmos} is an extension of the public code {\tt MontePython}\footnote{\url{ https://github.com/baudren/montepython_public}} \citep{Audren2013}. We compute evidence ratios, information and Suspiciousness using the public code {\tt anesthetic}\footnote{\url{https://github.com/williamjameshandley/anesthetic}} \citep{anesthetic}. We first reproduce the results from Table 2 of K19\footnote{We use natural logarithms, while K19 use base $10$ logarithms.} ($\log R = 4.21 \pm 0.15$), and then apply our method to the same split of the data. We get a log Suspiciousness $\log S = -1.992 \pm 0.064$. Under the Gaussian approximation, we can assign a {\it tension probability} $p_t$  to this value: we get the number of constrained dimensions $n_d = 2.75 \pm 0.18$, and the corresponding tension probability of the data sets being consistent is $p = 0.0674 \pm 0.0062$, corresponding to a significance $1.83 \sigma$. 

As expected, in this case the Suspiciousness provides a far more accurate assessment of tension than does $R$, which obtains `very strong' evidence for agreement according to the Jeffreys scale used in K19. However, the significance is also lower than in E18, which is a reflection on the dependence on overall goodness of fit of the statistic used in that paper. In other words, our method quantifies internal consistency of the data, but is `blind' to the goodness of fit, while the method used in E18 was sensitive to both. This is a very desirable quality in an internal consistency test, as it allows us to detect the origin of a potential problem in the data. 
For the KiDS-450 data, the Suspiciousness says that bin 3 is on the edge between consistent and moderately inconsistent, because it measures only the degree of agreement between the predictions of this bin and the rest of the data. However, we stress the need to combine a Suspiciousness internal consistency test with a reliable test of overall goodness of fit: the Suspiciousness alone would have been insufficient to identify the suboptimal KiDS-450 covariance matrix, while the E18 tests did so precisely because of their sensitivity to the overall goodness of fit. 

As a final test of our statistical method, we repeat the sensitivity analysis of appendix A1 in K19: instead of using the real KiDS-450 data, we use mock data vectors, in which the source redshift distribution is shifted by $dz_3 = {0, 0.15, 0.20}$. Our results are shown in \cref{fig:sens} and \cref{table:sens}. Again, we see how the Bayes ratio $R$ can hide tensions because of the width of the priors, while the Suspiciousness shows clear tension for both shifts of the redshift distribution. For example, $dz_3 = 0.15$ corresponds to `No Evidence' in $R$, while for the same shift $S$ yields a more than $3 \sigma$ discrepancy, indicating a strong tension. This highlights the value of the Suspiciousness for internal consistency tests in large data sets. The table also shows how the number of sigma is zero for $dz_3 = 0$. This is because in this case we are comparing data sets whose posteriors overlap perfectly. We should generally also be suspicious of cases where the tension probability is very close to one (such as this one), as they indicate that the agreement is `too good'. 


\vspace{-15pt}
\section{Conclusions}
\label{sec:conc}

In this paper, we extended the novel tension metric introduced in \cite{Handley2019a} to the case of correlated data sets. This takes the role of ``tier 1'' consistency tests introduced in \cite{Kohlinger2019}, and provides a measure of consistency for correlated data sets that uses Bayesian quantities, is parameterization independent, and can be used for the case of wide, uninformative priors since it does not depend on the prior volume (in contrast to the Bayes ratio for data set comparison $R$ introduced in \citealt{Marshall2006}).

We applied this formalism to a Gaussian toy model, and used this to compare it with tiers two and three of K19, showing that we obtain the same results for Gaussian posteriors. We propose our method as a diagnostic tool of consistency between correlated data sets, that can be complemented by best-fit calculations (tier 2), and Posterior Predictive Distributions (tier 3) to identify the origin of existing internal tension. 

We applied this formalism to the case of KiDS-450 data from \cite{Hildebrandt2017}, focusing on tests of tomographic redshift bin 3 vs the rest of the data (bins 1,2 and 4). We find that the $2.60 \sigma$ tension detected in \cite{Efstathiou2018} reduces to $1.83 \sigma$. We interpret this as a difference in the tension metric used: the methods used in \cite{Efstathiou2018} depend on overall goodness of fit, while those used in this work do not. This is a desirable quality, as it allows us to identify the origin of potential problems in the data, but it stresses the need to combine the Suspiciousness with a goodness of fit test. 

This method can generally be used for correlated data sets where uninformative priors are used. In particular, it can be used for internal consistency tests of cosmological surveys (such as DES and KiDS), and also to quantify tension between data sets with a non-negligible covariance (such as {\it Euclid} and LSST). We conclude that due to its Bayesian nature and intuitive interpretation, this methods serves as a perfect diagnostic of internal consistency, and we encourage present and future cosmological surveys to use it.

\section*{Acknowledgements}

Based on data products from observations made with ESO Telescopes at the La Silla Paranal Observatory under programme IDs 177.A-3016, 177.A-3017 and 177.A-3018, and on data products produced by Target/OmegaCEN, INAF- OACN, INAF-OAPD and the KiDS production team, on behalf of the KiDS consortium. We would like to thank Catherine Heymans for useful comments. 

PL \& OL acknowledge STFC Consolidated Grant ST/R000476/1. FK acknowledges support from the World Premier International Research Center Initiative (WPI), MEXT, Japan. WJH thanks Gonville \& Caius College for their continuing support via a Research Fellowship.


\section*{Appendix}
\label{sec:app}

We use \cref{eq:corr} to derive \cref{eq:sep} (with the associated definitions in \cref{eq:system0} through \cref{eq:system3}) using the general `complete the square' formula
\begin{multline}
\label{eq:general_complete_the_square}
(Q \theta - K)^\T \Sigma^{-1} (Q \theta - K) = \\
(\theta - C)^\T (Q^\T \Sigma^{-1} Q) (\theta - C) + K^\T \Sigma^{-1} (K - Q C)
\end{multline}
where $C = (Q^\T \Sigma^{-1} Q)^{-1} Q^\T \Sigma^{-1} K$. In \cref{eq:general_complete_the_square} set
\begin{equation}
Q = {\begin{bmatrix}I \\ I\end{bmatrix}}, K = {\begin{bmatrix}\mu_A \\ \mu_B\end{bmatrix}}  \text{ and } \Sigma = \Sigma_1 \text{ (so } \Sigma^{-1} = \Lambda \text{)}. 
\end{equation}
Also define $P = {\begin{bmatrix}I & -I\end{bmatrix}}$ (so $PK = \mu_A - \mu_B$) and set
\begin{equation}
M^{-1} = \Sigma_A - \Sigma_X - \Sigma_X^\T + \Sigma_B = P \Sigma P^\T
\end{equation}
and
\begin{equation}
L^{-1} = \Lambda_A + \Lambda_X + \Lambda_X^\T + \Lambda_B = Q^\T \Lambda Q.
\end{equation}
The covariance matrix in the RHS of \cref{eq:general_complete_the_square} is
\begin{equation}
\label{eq:appendix_covariance}
Q^\T \Sigma_1^{-1} Q = Q^\T \Lambda Q = L^{-1};
\end{equation}
this will be $\Sigma_0^{-1}$ in \cref{eq:sep}.
For the constant term on the RHS of \cref{eq:general_complete_the_square}, begin with
\begin{equation}
\begin{split}
0 = PQ = (P \Sigma) (\Lambda Q) &= (\Sigma_A - \Sigma_X^\T)(\Lambda_A + \Lambda_X) +\\
&\qquad \qquad (\Sigma_X - \Sigma_B)(\Lambda_X^\T + \Lambda_B) \text{, so}\\
(\Lambda_B + \Lambda_X^\T)(\Lambda_A + \Lambda_X)^{-1} &= (\Sigma_B - \Sigma_X)^{-1}(\Sigma_A-\Sigma_X^\T)\\
L^{-1}(\Lambda_A + \Lambda_X)^{-1} - I &= (\Sigma_B - \Sigma_X)^{-1} M^{-1} - I\\
(\Lambda_A + \Lambda_X)L &= M(\Sigma_B - \Sigma_X).
\end{split}
\end{equation}
This establishes the upper-right block of the identity $\Lambda Q L Q^\T + P^\T M P \Sigma_1 = I$ and the other blocks can be established similarly. Thus
\begin{equation}
\label{eq:appendix_constant_term}
\begin{split}
K^\T \Sigma_1^{-1}(K-QC) &= K^\T (I-\Lambda Q L Q^\T) \Lambda K\\
&=(P K)^\T M P K = (\mu_A-\mu_B)^\T \Sigma_{\Delta \mu}^{-1} (\mu_A-\mu_B);
\end{split}
\end{equation}
this will contribute to $\log \Lmax_0$ in \cref{eq:sep}. Finally the mean used in the RHS of \cref{eq:general_complete_the_square} is
\begin{equation}
\begin{split}
\label{eq:appendix_mean}
C &= (Q^\T \Sigma^{-1} Q)^{-1} Q^\T \Sigma^{-1} K = L Q^\T \Sigma_1^{-1} K = \Sigma_0 Q^\T \Lambda K\\
&= \Sigma_0 \left((\Lambda_A + \Lambda_X^\T) \mu_A + (\Lambda_X + \Lambda_B) \mu_B \right);
\end{split}
\end{equation}
this will be $\mu_0$ in \cref{eq:sep}.

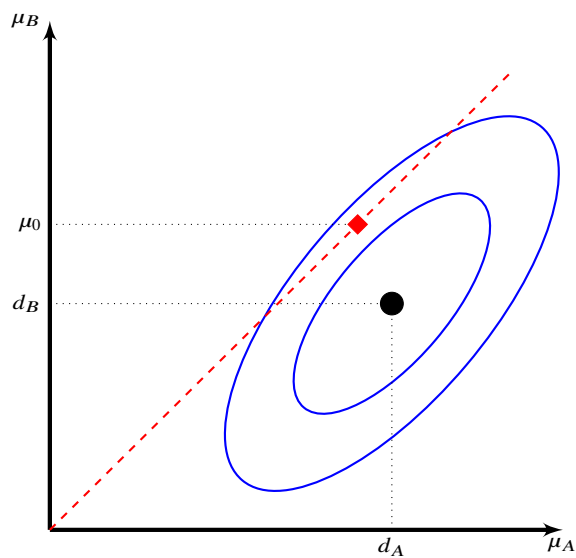
\begin{figure}
\centering
\begin{tikzpicture}[scale=1.5,>=latex']

    \def\graphlimit{4.5}
    \draw [ultra thick, ->] (0, 0) -- (\graphlimit,0);
    \node [below] at (\graphlimit,0) {$\mu_A$};
    \draw [ultra thick, ->] (0, 0) -- (0,\graphlimit);
    \node [left] at (0,\graphlimit) {$\mu_B$};

    \def\a{3.0}
    \def\b{2.0}
    \def\mean{(\a,\b)}
    
    \def\angle{-40} 
    \def\cosineofangle{0.766044443}
    \def\sineofangle{-0.64278761}
    
    \def\rotatedmean{(\cosineofangle*\a+\sineofangle*\b,-\sineofangle*\a+\cosineofangle*\b)}

    \def\Pi{\rotatedmean ellipse (.5 and 1.2)}
    \def\Pii{\rotatedmean ellipse (.85 and 2.04)}
    \tikzstyle{P_1} = [draw,blue,thick]
    \begin{scope}[rotate=\angle]
        \path[P_1] \Pi;
        \path[P_1] \Pii;
    \end{scope}
    
    \draw[fill] \mean circle [radius=0.1];
    \draw [dotted, thin, black] (\a, 0) -- (\a, \b);
    \node [below] at (\a, 0) {$d_A$};
    \draw [dotted, thin, black] (0, \b) -- (\a, \b);
    \node [left] at (0, \b) {$d_B$};

    \draw [dashed,thick,red] (0,0) -- (\graphlimit*0.9,\graphlimit*0.9);
    
    \def\onedmode{2.7} 
    \draw [dotted, thin, black] (0, \onedmode) -- (\onedmode, \onedmode);
    \node [left] at (0, \onedmode) {$\mu_0$};
    \node[red,diamond,draw,fill,scale=0.65] at (\onedmode,\onedmode) {};

\end{tikzpicture}
\caption{Inference of mean(s) from two correlated data observations $d_A$ and $d_B$. The blue contours show (under hypothesis $H_1$ i.e. separate means) the likelihood (assumed Gaussian with covariance $\Sigma$) of the mean values $(\mu_A,\mu_B)$; the maximum likelihood (and with flat priors the mode of the posterior) for these means is the black circle at $(d_A, d_B)$. Under hypothesis $H_0$ (i.e. one common mean) the likelihood is restricted to the red dashed line $\mu_A=\mu_B$ where it will be Gaussian with covariance $\Sigma_0$; here the maximum likelihood (and hence posterior mode) is at the red diamond $(\mu_0,\mu_0)$.} \label{fig:appendix}

\end{figure}

See \cref{fig:appendix} for a diagram illustrating the behaviour of the likelihood function under the two hypotheses.

\vspace{-10pt}
\bibliographystyle{mnras}
\bibliography{refs}

\end{document}